\documentclass{article}[12pt]
\usepackage{cleveref}
\usepackage{color}
\usepackage{graphicx,amssymb,amsfonts,amsmath,amssymb,amscd,amstext}

\pdfoutput=1
\def\be{\begin{eqnarray}}
\def\ee{\end{eqnarray}}

\def\tr{\operatorname{tr}}

\def\Vol{\operatorname{Vol}}
\def\vol{\operatorname{vol}}
\def\tr{\operatorname{tr}}
\def\Id{\operatorname{Id}}

\begin{document}
\thispagestyle{empty}

\begin{flushright}
YITP-SB-14-20 
\end{flushright}

\begin{center}
\vspace{1cm} { \LARGE {\bf 
Universal Thermal Corrections to Entanglement Entropy for Conformal Field Theories on Spheres
}}
\vspace{1.1cm}

Christopher P. Herzog \\
\vspace{0.8cm}
{ \it C.~N.~Yang Institute for Theoretical Physics, 
Department of Physics and Astronomy \\
Stony Brook University, Stony Brook, NY  11794}

\vspace{0.8cm}

\end{center}

\begin{abstract}
\noindent
We consider entanglement entropy of a cap-like region for a  conformal field theory living on a sphere times a circle in $d$ space-time dimensions.  Assuming that the finite size of the system introduces a unique ground state with a nonzero mass gap, we calculate the leading correction to the entanglement entropy in a low temperature expansion.  The correction has a universal form for any conformal field theory that depends only on the size of the mass gap,  its degeneracy, and the angular size of the cap.  We confirm our result by calculating the entanglement entropy of a conformally coupled scalar numerically.  We argue that an apparent discrepancy for the scalar can be explained away through a careful treatment of boundary terms.
In an appendix, to confirm the accuracy of the numerics, we study the mutual information of two cap-like regions at zero temperature.

\end{abstract}

\pagebreak
\setcounter{page}{1}

\section{Introduction}

Studies of entanglement entropy sit at a nexus of many different areas of theoretical physics.  Central to quantum information, communication, and computation, entanglement entropy can also be used to  detect exotic phase transitions in many-body systems  lacking a local order parameter \cite{Osborne:2002zz,Vidal:2002rm}.  Certain specific types of entanglement entropy order quantum field theories under renormalization group flow \cite{Casini:2006es,Casini:2012ei}.   Entanglement entropy is also a key concept in attempts to understand the microscopic origin of black hole entropy (see e.g. 
\cite{Bombelli:1986,Srednicki:1993im}).  

To define the entanglement entropy, partition the Hilbert space into pieces $A$ and complement $\bar A$.  
Typically (and hereafter in this letter) $A$ and $\bar A$ correspond to spatial regions.  
Not all quantum systems may allow for such a partition.
The reduced density matrix is defined as a partial trace of the full density matrix $\rho$ 
over the degrees of freedom in $\bar A$: 
\begin{eqnarray}
\rho_A \equiv \tr_{\bar A} \rho \ .
\end{eqnarray}
The entanglement entropy is then the von Neumann entropy of the reduced density matrix:
\be
S_A \equiv - \tr \rho_A \log \rho_A \ .
\label{EEdef}
\ee

In this letter, we are interested in the entanglement entropy at nonzero temperature $T = 1/\beta$.  
The initial density matrix takes the standard Boltzmann form
\be
\rho = \frac{e^{-\beta H}}{\tr( e^{-\beta H})} \ ,
\label{thermalrho}
\ee
where $H$ is the Hamiltonian. 
 For thermal states, $S_A$ is no longer a good measure of quantum entanglement.  The entanglement entropy is contaminated by the thermal entropy of region $A$ and in the high temperature limit, becomes dominated by it.  To reveal the quantum entanglement of a thermal system, one should subtract off the thermal contribution to $S_A$.  
 
Ref.\ \cite{Herzog:2012bw} conjectured that for any quantum system with a mass gap $m_{\rm gap}$,
such corrections should scale as $e^{-\beta m_{\rm gap}}$ when $\beta m_{\rm gap} \gg 1$.  With a couple of modest assumptions, 
this conjecture follows from writing (\ref{thermalrho}) as a Boltzmann sum over states.  
Ref.\ \cite{Cardy:2014jwa} provided the form of the coefficient of the $e^{-\beta m_{\rm gap}}$ Boltzmann factor in the case where the system was described by a two dimensional conformal field theory.\footnote{%
 See refs.\ \cite{Azeyanagi:2007bj,Herzog:2013py,Datta:2013hba} for related results involving free fermions and scalars.
}
In particular, for a CFT on a circle of circumference $L$ in the case where $A$ consists of a single interval of length 
$\ell$, the correction is
\begin{eqnarray}
\label{dSEcorr2}
\delta S_A = S_A(T) - S_A(0) &=& 2g \Delta \left[1 - \frac{\pi \ell}{L} \cot \left( \frac{ \pi \ell}{L} \right) \right] e^{-2 \pi \Delta \beta/ L} 
+ o(e^{-2 \pi \Delta \beta / L})\ ,
\end{eqnarray} 
where $\Delta$ is the smallest scaling dimension among the set of operators including the stress tensor and all primaries not equal to the identity and $g$ is their degeneracy.  (See also \cite{Chen:2014unl} for the specific case of the stress tensor.)
In order for this result to hold, the CFT needs to have a unique ground state separated from the first excited state by a nonzero mass gap (induced by the finite volume of the system).  

More generally, we consider a CFT on $S^1 \times S^{d-1}$ where the radius of $S^1$ is $\beta/2\pi$ and of $S^{d-1}$ is $R$.  If we define $A \subset S^{d-1}$ to be the cap-like region with polar angles $\theta < \theta_0$, then the thermal correction has the low temperature scaling form
\be
\delta S_A(T) =  g \Delta \, I_d (\theta_0) e^{-\beta \Delta  / R} + o(e^{-\beta \Delta/R}) \ ,
\label{mainresult}
\ee
where\footnote{%
 The volume of a unit sphere can be expressed in terms of a gamma function, $\Vol(S^{d-1}) = 2 \pi^{d/2} / \Gamma( d/2)$. 
}
\be
I_d(\theta_0) \equiv 2 \pi  \frac{\Vol(S^{d-2})}{\Vol(S^{d-1})}  \int_0^{\theta_0} \frac{\cos \theta - \cos \theta_0}{\sin \theta_0}
\sin^{d-2} \theta \, d\theta\ .
\label{Idtheta}
\ee
 In order for our result to be valid, we assume that the first excited state $|\psi \rangle$ can be created in radial quantization by a local operator $\psi(x)$ acting at the origin.  

Several comments are in order.
\begin{itemize}

\item
The new result (\ref{mainresult}) matches the earlier result (\ref{dSEcorr2}) in the case $d=2$ and $L = 2 \pi R$, as it should. 

\item
The ``odd'' part of $I_d(\theta_0)$
leads to an elegant result for $S_A(T) - S_{\bar A}(T)$ in the low temperature limit, namely
\be
S_{\bar A}(T) - S_A(T) = 2 \pi g \Delta \cot (\theta_0) \, e^{-\beta \Delta/R} + o(e^{-\beta \Delta/R}) \ .
\label{mainresultb}
\ee
From a Schmidt decomposition of the Hilbert space (see for example \cite{EislerPeschel}), it follows that $S_{\bar A}(0) = S_A(0)$.  However at nonzero
temperature, the two entanglement entropies are generically no longer equal.

\item
For $d$ a positive integer, $I_d(\theta)$ can be expressed as a finite sum of trigonometric functions.  
Indeed, $I_d(\theta)$ satisfies a recurrence relation: 
\be
I_d(\theta) = -  2 \pi \frac{\Vol(S^{d-2})}{\Vol(S^{d-1})}  \frac{\sin^{d-2}\theta}{(d-1)(d-2)} + I_{d-2} (\theta) \ .
\label{recrel}
\ee
We give some specific examples of $I_d(\theta)$ for small $d$ in the text.

\end{itemize}

To check our result (\ref{mainresult}), we compute the entanglement entropy of a conformally coupled scalar field in $d>2$ and we find a discrepancy.  The numerics agrees remarkably well with the analytic result provided we make the substitution $I_d(\theta_0) \to I_{d-2}(\theta_0)$ in (\ref{mainresult}).  In view of the recurrence relation (\ref{recrel}), this discrepancy is proportional to $\sin^{d-2} \theta_0$ and thus also proportional to the area of $\partial A$.
To make sure that the numerics are functioning properly, we also study the mutual information of the conformally coupled scalar, which should be insensitive to such an area dependent discrepancy.  We are able to confirm some nontrivial, $d$ dependent predictions of refs.\ \cite{Shiba:2012np,Cardy:2013nua,Casini:2009sr}.

We believe the discrepancy is due to a subtle problem with the way our main result (\ref{mainresult}) was derived.  We make use of a conformal map from hyperbolic space to the region $A$ on the sphere.  The action for a conformally coupled scalar has a boundary term.  
However, the boundary of hyperbolic space 
used in the computation of eq.\ (\ref{mainresult}) is slightly different from the pull-back of $\partial A$ needed for the entanglement entropy calculation.  This difference can precisely account for the $\sin^{d-2} \theta_0$ discrepancy.
More generally for the $\sin^{d-2} \theta_0$ term in our main result (\ref{mainresult}) to be accurate, the conformal field theory needs to be insensitive to the differences between the two boundaries in question.
We leave fuller discussion of these issues to the text.  Note that the quantity $S_A(T) - S_{\bar A}(T)$ is independent of this $\sin^{d-2} \theta_0$ ambiguity.

%
%

The outline of the paper is as follows. In section \ref{sec:analytic}, we discuss the derivation of our main result (\ref{mainresult}).  
In section \ref{sec:numeric}, we confirm eq.\ (\ref{mainresult}) numerically for the case of a conformally coupled scalar field, up to the $\sin^{d-2} \theta_0$ discrepancy.
In section \ref{sec:discrepancy}, we show the discrepancy occurs because of a boundary term in the action for a conformally coupled scalar.  
In section \ref{sec:discussion}, we discuss some implications of our results and some areas for future work.
An appendix contains the studies of the mutual information.

\section{Analytical Calculation}
\label{sec:analytic}

We are interested in a $d$ dimensional CFT on $S^{d-1}$ at finite temperature.  We will assume that the $S^{d-1}$ gaps the spectrum and leads to a unique ground state.  (Maximally supersymmetric Yang-Mills in 3+1 dimensions on an $S^3$ would be an example.)
We write down the density matrix as a Boltzmann sum, keeping only the ground state $|0 \rangle$ and the first excited states $| \psi_i \rangle$ with $i = 1, \ldots, g$:
\be
\rho = \frac{|0 \rangle \langle 0 | + 
\sum_i | \psi_i \rangle \langle \psi_i | e^{-\beta E_\psi} + \ldots}{1 + g e^{-\beta E_\psi} + \ldots}
\ee
Consider a cap-like region $A$ that extends from the north pole of the $S^{d-1}$ down to a latitude $\theta_0$ and its complement $\bar A$. We would like to compute the leading order change in the entanglement entropy of region $A$ due to temperature:
\be
S_A(T) - S_A(0) \equiv \delta S_A = \sum_i \tr \left[ \left( \tr_{\bar A} |\psi_i \rangle \langle \psi_i | -  \tr_{\bar A} |0 \rangle \langle 0 | \right) H_M \right] e^{-\beta E_\psi} + \ldots\ ,
\label{dSAinitial}
\ee
where $H_M \equiv - \log \tr_{\bar A} |0 \rangle \langle 0 |$.
For a conformal field theory on ${\mathbb R} \times S^{d-1}$, $E_\psi = \Delta / R$ where $\Delta$ is the scaling dimension of the operator that created the degenerate states $|\psi_i \rangle$ and $R$ is the radius of the sphere.  

The next step in the argument makes heavy use of results from ref.\ \cite{Casini:2011kv}.\footnote{%
 We thank 
 H.~Casini and 
 N.~Lashkari for 
 drawing our attention to these results.
}
 The point is that through a Weyl scaling and coordinate redefinition, the modular Hamiltonian $H_M$ can be expressed as an integral over the $tt$ component of the stress tensor on ${\mathbb R} \times S^{d-1}$.
We use the conformal transformation between ${\mathbb R} \times S^{d-1}$ and ${\mathbb R} \times H^{d-1}$ (where $H^d$ is $d$ dimensional hyperbolic space) described in section 2.3 of 
\cite{Casini:2011kv}:
\begin{eqnarray}
ds^2 &=& -dt^2 + R^2 (d\theta^2 + \sin^2 \theta \, d\Omega_{d-2}^2) \\
 &=& \Omega^2 [ - d\tau^2 + R^2 (du^2 + \sinh^2 u \, d \Omega_{d-2}^2)] \ ,
\end{eqnarray}
where $d\Omega_d^2$ is a line element on a unit $S^{d}$,
\begin{eqnarray}
\tan(t/R) &=& \frac{ \sin \theta_0 \sinh(\tau/R)}{\cosh u + \cos \theta_0 \cosh(\tau/R)} \ , \\
\tan \theta &=& \frac{\sin \theta_0 \sinh u}{\cos \theta_0 \cosh u + \cosh (\tau/R)} \ ,
\end{eqnarray}
and $\Omega \sinh u = \sin \theta$.
 This map takes all of $H^{d-1}$ at $\tau=0$ to the region $A \subset S^{d-1}$ at $t=0$.  

On ${\mathbb R} \times H^{d-1}$, the claim is that the modular Hamiltonian $H_M$ is an integral 
of $T_{\tau\tau}$ over the volume of $H^{d-1}$ at $\tau=0$:
\be
H_M = 2 \pi R^d \int_{0}^\infty   \int_{S^{d-2}}  T_{\tau \tau}(q)   \vol(S^{d-2})  (\sinh u)^{d-2} \, du \ ,
\ee
where $\vol(S^{d})$ is short hand for the volume form on $S^{d}$ and $q \in H^{d-1}$ is a point in hyperbolic space.
(We will argue in section \ref{sec:discrepancy} that at least for a conformally coupled scalar, 
this integral may differ from the true modular Hamiltonian by boundary terms.)
The covariance of a CFT under Weyl rescaling allows us to rewrite $H_M$ in terms of $T_{tt}$ on 
${\mathbb R} \times S^{d-1}$.  Note that at $\tau = 0$, $\partial \theta / \partial \tau$ vanishes.
It follows then that at $\tau=0$,
\be
T_{\tau\tau} = \Omega^{d-2} \left( \frac{\partial t}{\partial \tau} \right)^2 T_{tt} + \ldots \ .
\ee
We can ignore the Schwarzian derivative contribution, indicated by the ellipsis, because the density matrices are normalized to one.
To express $H_M$ in terms of quantities on ${\mathbb R} \times S^{d-1}$, we note 
that
\be
\left. \frac{\partial u}{\partial \theta} \left( \frac{\partial t}{\partial \tau} \right)^2\right|_{\tau=0} = \frac{\cos \theta - \cos \theta_0}{\sin \theta_0} \ .
\ee
Assembling the pieces, we obtain
\begin{eqnarray}
H_M = 2\pi R^d \int_0^{\theta_0} \int_{S^{d-2}} T_{tt}(p) \frac{\cos \theta - \cos \theta_0}{\sin \theta_0} 
\sin^{d-2} \theta \,\vol(S^{d-2})   d \theta \ ,
\label{HMresult}
\end{eqnarray}
where $p \in A$ is a point.
(This calculation of $H_M$ in $d=2$ was carried out in ref.\ \cite{Blanco:2013joa}.)

It remains to evaluate the trace
\be
I_A \equiv \sum_i \tr \left[ \left(\tr_{\bar A} |\psi_i \rangle \langle \psi_i | - \tr_{\bar A} | 0 \rangle \langle 0 | \right) T_{tt}(p) \right]  \ ,
\label{IAinitial}
\ee
for an arbitrary point $p \in A$.  By locality, this trace cannot be affected by the partial traces over region $\bar A$ and one finds that $I_A$ is the local difference in energy density between the ground and first excited states:
\be
I_A = \sum_i \left( \langle \psi_i | T_{tt}(p) | \psi_i \rangle - \langle 0 | T_{tt}(p) | 0 \rangle \right) \ .
\ee
The states $|\psi_i \rangle$ must transform under some representation of $SO(d)$ because of the rotational symmetry of the sphere while the operator $T_{tt}(p)$ will transform as a scalar under rotations.  
It follows that $I_A$ must also transform as a scalar.  In fact, having summed over $i$, by rotational symmetry, there is no longer any way for $I_A$ to be angle dependent;  $I_A$ can only be the constant function.  Because the states $|\psi_i \rangle$ are normalized to one, the integral over a single state gives the energy of that state:
\be
\int_{S^{d-1}} \left( \langle \psi_i | T_{tt}(p) | \psi_i \rangle- \langle 0| T_{tt}(p) | 0 \rangle  \right) \vol(S^{d-1})  = \Delta / R \ .
\ee
It follows then that $I_A$ is the constant energy density associated with the mass gap\footnote{%
 This argument appears to be a generalization of Uns\"old's Theorem in quantum mechanics, that $4 \pi \sum_{m} |Y_{lm}|^2 = 2l+1$.
}
\be
I_A = g \frac{\Delta}{R^d \Vol(S^{d-1})} \ .
\label{IAresult}
\ee
Alternately, one can look at the precise form of the three point function $\langle \psi_i(p_1) T_{tt}(p) \psi_i(p_2) \rangle$ 
(see ref.\ \cite{Osborn:1993cr} for example), a procedure which becomes cumbersome for higher spin operators.

Our main result (\ref{mainresult}) 
\[
\delta S_A =  g \Delta \,  I_d (\theta_0) e^{-\beta \Delta  / R}  + \ldots \ .
\]
now follows directly from (\ref{dSAinitial}), (\ref{HMresult}), (\ref{IAinitial}), and (\ref{IAresult}).  The integral $I_d(\theta_0)$ 
was defined in (\ref{Idtheta}).
Starting from the integral definition, one can deduce the 
recurrence relation (\ref{recrel}) mentioned in the introduction.
In our numerical calculation for the conformally coupled scalar in $d>2$, we will see no $\sin^{d-2} \theta_0$ dependence at all in $\delta S_A$.   The numerics yields the result (\ref{mainresult}) but where $I_d(\theta_0)$ is replaced with $I_{d-2}(\theta_0)$.  
As we explain in section \ref{sec:discrepancy}, the discrepancy is caused by a boundary term in the action for the conformally coupled scalar.  

For some small dimensions, we obtain\footnote{%
 Although the integral diverges, the expression for $I_1$ can at least be defined formally from the recurrence relation.
}
\begin{eqnarray}
I_1(\theta) &=& \pi \tan \frac{\theta}{2} \ , \; \; \;
I_2(\theta) = 2 (1 - \theta \cot \theta) \ , \\
I_3(\theta) &=& 2 \pi \csc \theta \sin^4 \frac{\theta}{2} \ , \; \; \;
I_4(\theta) = \frac{1}{3} (5 + \cos 2 \theta - 6 \theta \cot \theta) \ ,\\
I_5(\theta) &=& \frac{\pi}{2} (3 + \cos \theta) \sin^4 \left( \frac{\theta}{2} \right) \tan \left( \frac{\theta}{2} \right)\ , \\
I_6(\theta) &=& \frac{16}{15} \sin^4 \theta - \frac{1}{6} ( 12 \theta - 8 \sin 2 \theta + \sin 4 \theta) \cot \theta\ .
\end{eqnarray}
We also find the following simple form for the ``odd'' part:
\be
I_d(\pi - \theta_0) - I_d(\theta_0) =2 \pi \frac{\Vol(S^{d-2})}{\Vol(S^{d-1})}  \int_0^\pi  \frac{\cos \theta_0 - \cos \theta}{\sin \theta_0} \sin^{d-2} \theta \, d\theta = 2 \pi \cot(\theta_0) \ ,
\ee
which leads to the $d$ independent result (\ref{mainresultb}) discussed in the introduction.  Note that since $\sin \theta$ is invariant under $\theta \to \pi -\theta$, the odd part is insensitive to potential ambiguities in the $\sin^{d-2} \theta_0$ term.

\section{Conformally Coupled Scalar}
\label{sec:numeric}

\begin{figure}
 \begin{center}
  \includegraphics[width=2.4in]{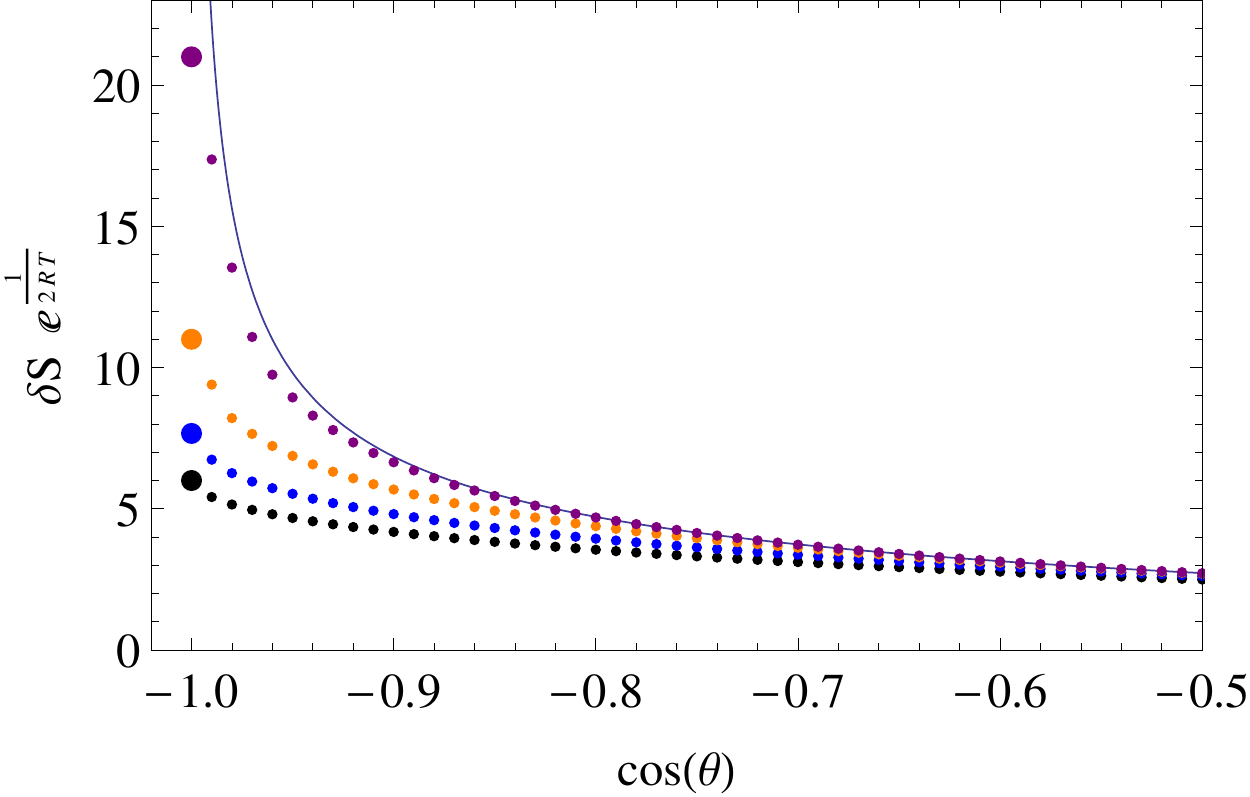}
 \includegraphics[width=2.3in]{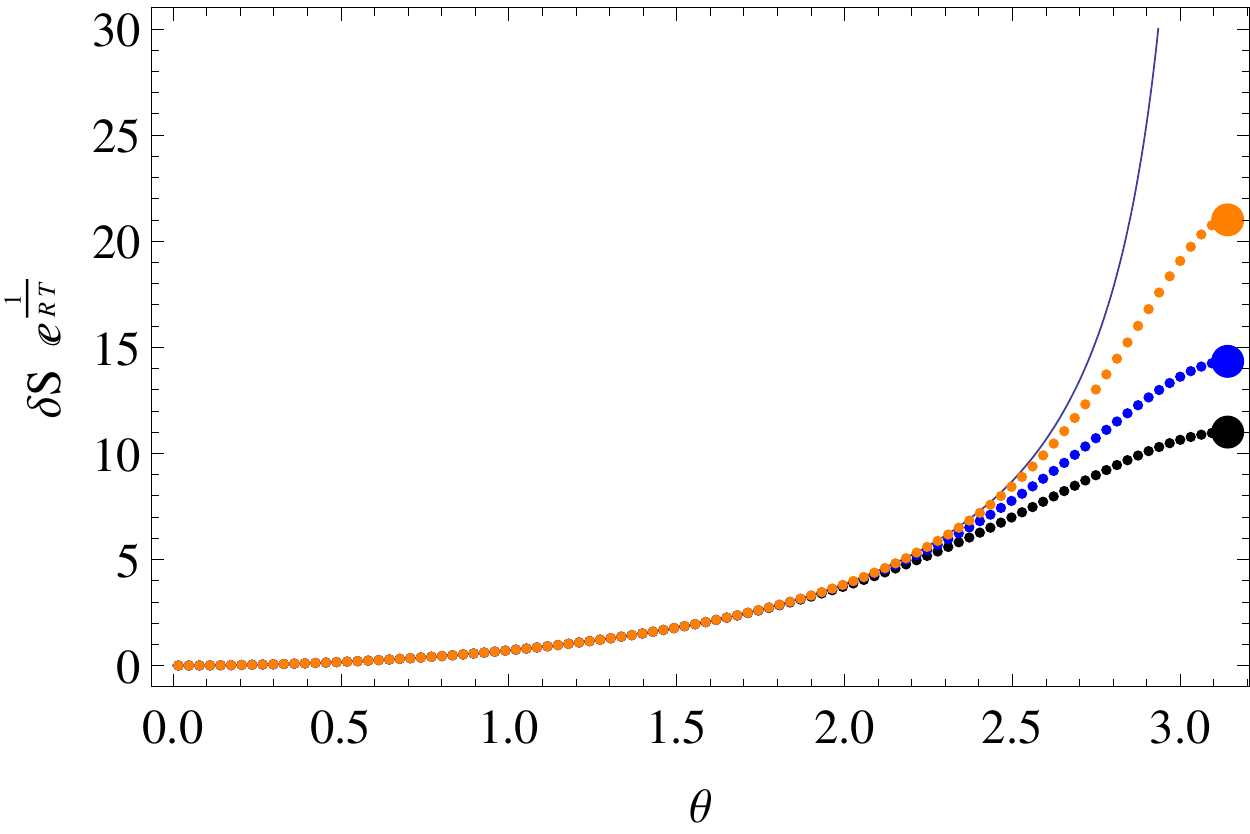}
   \end{center}
  \caption{ The entanglement entropy difference $\delta S = S_0(T) - S_0(0)$ for a caplike region with angular size $\theta$.  The plot demonstrates the cross over between small $T$ and small $\pi - \theta$ behavior.
  Left: Three dimensional case. From top to bottom, the data points correspond to $RT = 0.025$, 0.05, 0.075, and 0.1. Right: Four dimensional case. From top to bottom, the data points correspond to $RT = 0.05$, 0.075, and 0.1.  The curves are the prediction (\ref{mainresult}) with $I_d(\theta)$ replaced by $I_{d-2}(\theta)$, as discussed in the text.   The big dots mark the low temperature thermal entropy correction $1 + \Delta/RT$.  The lattice used had 200 grid points.
   }
\label{deltaSconstantT}
 \end{figure}

\begin{figure}
 \begin{center}
   \includegraphics[width=2.3in]{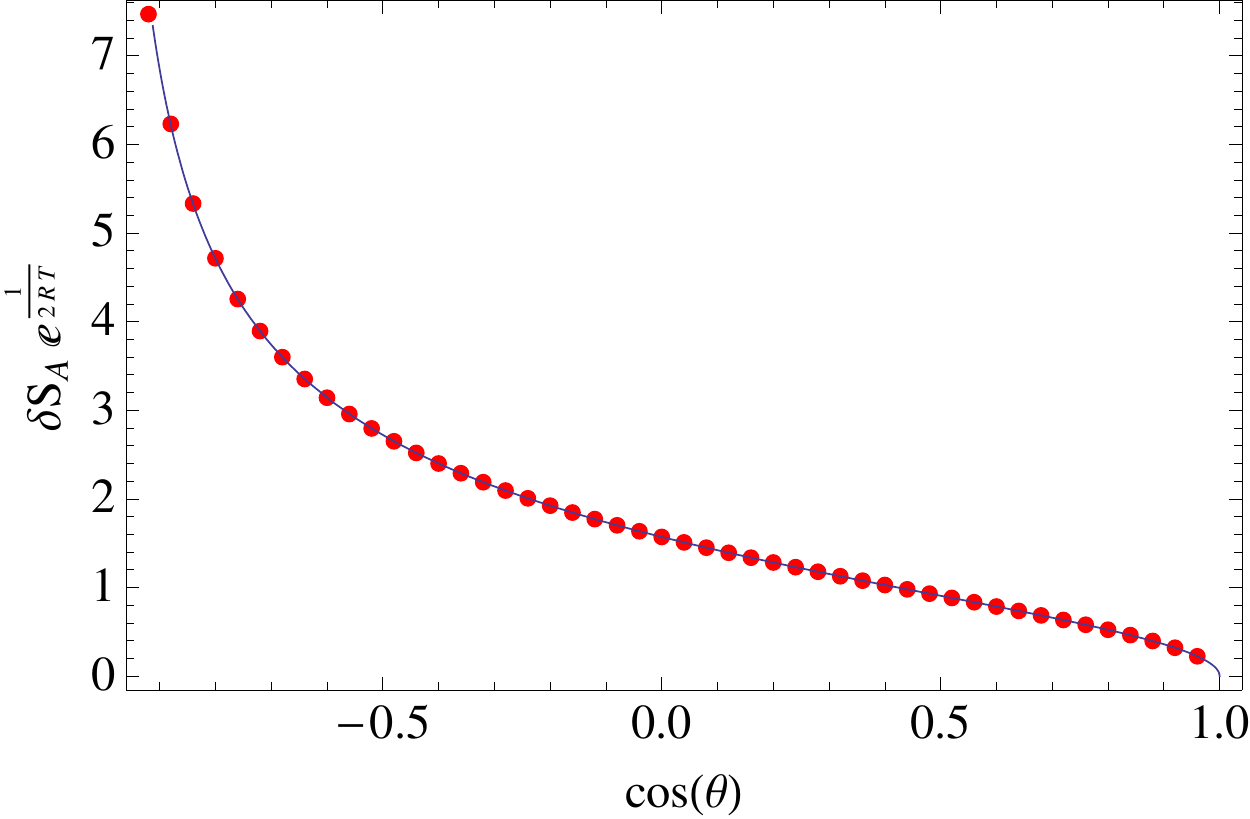}
  \includegraphics[width=2.37in]{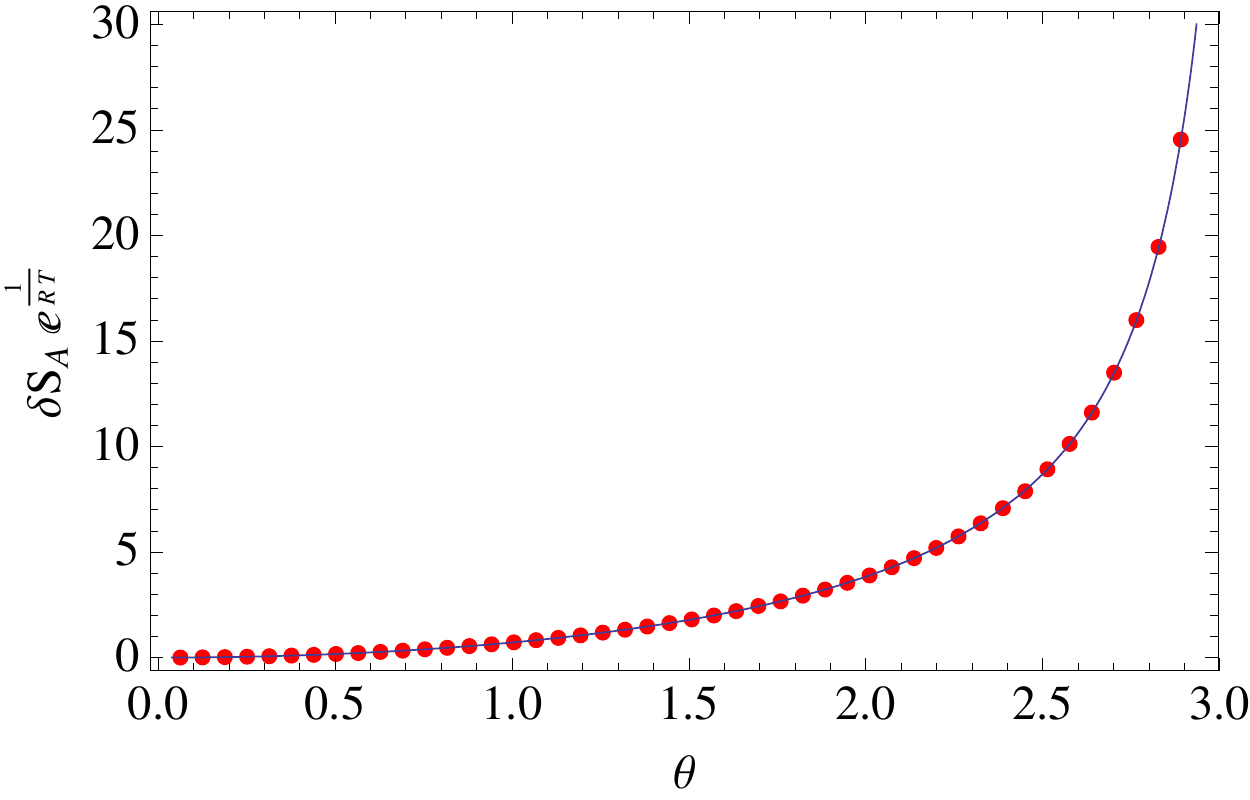}
   \end{center}
  \caption{ The entanglement entropy difference $\delta S_A = S_A(T) - S_A(0)$ for a cap like region $A$ with angular size $\theta$ in the limit where $T$ is sent to zero first: (left) $d=3$; (right) $d=4$.   The curves are the prediction (\ref{mainresult}) with $I_d(\theta)$ replaced by $I_{d-2}(\theta)$, as discussed in the text. The points were numerically determined.  The lattice used had 200 grid points.
     }
\label{deltaSTfirst}
 \end{figure}

We check our result (\ref{mainresult}) numerically by considering the case of a free, conformally coupled scalar field:
\be
S = -\frac{1}{2} \int d^d x \sqrt{-g} [ (\partial_\mu \phi) (\partial^\mu \phi) + \xi {\cal R} \phi^2 ] \ ,
\ee
where the conformal coupling is 
\[
\xi = \frac{d-2}{4(d-1)} \ .
\] 
For the manifold ${\mathbb R} \times S^{d-1}$, we write the line element as
\be
ds^2 = - dt^2 + R^2 (d \theta^2 + \sin^2 \theta \, h_{ab} \, d\theta^a d\theta^b) \ .
\ee
We obtain the Ricci scalar and effective mass:
\[
{\cal R} = \frac{(d-1)(d-2)}{R^2} \Longrightarrow m_{\rm eff}^2 = \left( \frac{d-2}{2R} \right)^2 \ ,
\]
where $R$ is the radius of the $S^{d-1}$.

The method we use is a modernized and slighly altered
version of the method described in \cite{Srednicki:1993im}.  The discretized free scalar is a collection of coupled harmonic oscillators.  The idea behind the method is write down the density matrix as a Gaussian integral and to perform the trace over its eigenvalues explicitly.  Following \cite{Herzog:2012bw,EislerPeschel,Casini:2009sr}, we reformulate the problem in terms of field $\phi$ and conjugate momentum
$\pi = \partial_t \phi$ two point functions.

To fix notation, we review how to quantize the scalar on the sphere $S^{d-1}$.
In canonical quantization, we must enforce the following commutation relation
\be
[ \phi(t,x), \pi(t,x') ] \sqrt{-g} = i \, \delta(x-x') \ .
\ee
From the canonical momentum, we may construct the Hamiltonian density\footnote{%
 One may construct an alternate energy density from the stress tensor,
 obtained by varying the action with respect to the metric.  The two densities differ by a well-known ``improvement'' 
 term, in this case proportional to $\vec \nabla^2 \phi^2$, where $\vec \nabla^2$ is the Laplacian on $S^{d-1}$.  One may worry that our
 entanglement entropy calculation is sensitive to this choice.  However, 
 the improvement term is a total derivative, and, because $S^{d-1}$ has no boundary, does not affect the Hamiltonian itself.
}
\begin{eqnarray}
 {\cal H} 
&=&\frac{(R \sin \theta)^{d-2} \sqrt{h}}{2R}\left\{ R^2 \pi^2 
 +  
(\partial_\theta \phi)^2 + \frac{h^{ab}(\partial_a \phi) (\partial_b \phi) }{\sin^2 \theta} + \frac{(d-2)^2}{4 } \phi^2  \right\} \ .
\end{eqnarray}
We write
\begin{eqnarray}
\phi &=& \sum_{\vec l } \Phi_{\vec l} (\theta) Y_{\vec l}(\theta_1, \ldots, \theta_{d-2}) \sqrt{R^{1-d} \sin^{2-d} \theta} \ , \\
\pi &=& \sum_{\vec l} \Pi_{\vec l} ( \theta) Y_{\vec l} (\theta_1, \ldots, \theta_{d-2})   \sqrt{R^{1-d} \sin^{2-d} \theta} \ ,
\end{eqnarray}
where $Y_{l_1, \ldots, l_{d-2}}(\theta_1, \ldots, \theta_{d-2})$ is a generalized (real)
spherical harmonic with $|l_1| \leq l_2 \leq \cdots \leq l_{d-2}$ and $\Delta_{S^{d-2}} Y_{l_1, \ldots, l_{d-2}} = - l_{d-2} (l_{d-2} +d-3)$.
It follows that
\[
[ \Phi_{\vec l}(\theta), \Pi_{\vec l'} (\theta')]  = i \delta_{\vec l, \vec l'}   \delta(\theta - \theta') \ .
\]
We write the Hamiltonian as $H = \sum_{\vec l} H_{\vec l}$. 
From the normalization and the definition of the $Y_{l_1, \ldots, l_{d-2}}(\theta_1, \ldots, \theta_{d-2})$, each term in the sum over angular modes can be written
\begin{eqnarray}
\label{Hvecl}
H_{\vec l} 
&=& \frac{1}{2R^2} \int_0^\pi  \biggl\{ R^2 \Pi_{\vec l}^2 
%
- \Phi_{\vec l} \partial_\theta^2 \Phi_{\vec l} + \frac{1}{4}(2m+d-2)(2m+d-4) \frac{\Phi_{\vec l}^2}{\sin^2 \theta} \biggr\} d\theta \ ,
 \end{eqnarray}
where we set $m \equiv l_{d-2}$ and dropped a total derivative.  
The number of spherical harmonics with $l_{d-2} = m$ is
\be
\operatorname{dim}(m) = {d+m -2 \choose d-2} - {d+m-4 \choose d-2} \ .
\ee

We can diagonalize this $H_{\vec l}$ using an orthogonal transformation involving associated Legendre functions: $\Phi_{\vec l} = \sum_l U_l(\theta) \tilde \Phi_l$ and $\Pi_{\vec l} = \sum_l U_l(\theta) \tilde \Pi_l$.  More specifically
\be
\Phi_{\vec l} (\theta) = \sum_{l=m}^\infty {\cal N}_{l,m}  \cdot \sqrt{\sin \theta} \, P^{-m-(d-3)/2}_{l + (d-3)/2}(\cos \theta) \cdot \tilde \Phi_l \ ,
\label{contlimit}
\ee
where the normalization factor is
\be
{\cal N}_{l,m} = \sqrt{ \frac{2l+d-2}{2} \frac{(l+m+d-3)!}{(l-m)!}} \ .
\ee
We can then write the Hamiltonian as a number of decoupled harmonic oscillators
\be
H_{\vec l} = \frac{1}{2} \sum_{l=m}^\infty \left\{ \tilde \Pi_{l}^2  + 
\omega_l^2 \tilde \Phi_l^2 \right\} \ ,
\ee
with mass equal to one and frequency
\be
\omega_l = \frac{1}{R} \left( l + \frac{d-2}{2} \right) \ .
\ee

In the continuum, the thermal two point functions from which we can reconstruct the entanglement entropy are then
\begin{eqnarray}
\label{phiphi}
\langle \Phi_{\vec l} (\theta) \Phi_{\vec l} (\theta') \rangle &=& \frac{1}{2} \sum_{l=m}^\infty U_l(\theta) \frac{1}{ \omega_l} \coth \frac{\omega_l}{2T} U_l(\theta') \ , \\
\label{pipi}
\langle \Pi_{\vec l} (\theta) \Pi_{\vec l} (\theta') \rangle &=& \frac{1}{2} \sum_{l=m}^\infty U_l(\theta)  \omega_l \coth \frac{\omega_l}{2T} U_l(\theta') \ .
\end{eqnarray}
We define the matrix $C_m(\theta_1, \theta_2)$ such that
\be
C_m(\theta_1, \theta_2)^2 = \int_0^{\theta_0} \langle \Phi_{\vec l} (\theta_1) \Phi_{\vec l} (\theta) \rangle
\langle \Pi_{\vec l} (\theta) \Pi_{\vec l} (\theta_2) \rangle d \theta \ .
\ee
The range of $C_m$ is restricted such that $0 \leq \theta_i \leq \theta_0$, $i=1$,2.   
The entanglement entropy contribution from $H_{\vec l}$ to $S_A$ is then
\be
S_m = \tr \left[ \left( C_m + \frac{1}{2} \right) \log \left( C_m + \frac{1}{2} \right) - \left(C_m - \frac{1}{2} \right) \log \left(C_m - \frac{1}{2} \right) \right] \ .
\ee
The entanglement entropy of region $A$ is the sum
\be
S_A = S_0 + \sum_{m=1}^\infty \dim(m) S_m \ . 
\label{SAfreedef}
\ee
In general, this infinite sum on $m$ needs to be treated with care.  However, in our particular case, we are interested in a low temperature limit.  In the difference $S_A(T) - S_A(0)$, the contributions from $m > 0$ are exponentially suppressed compared with $m=0$.

While the two point functions (\ref{phiphi}) and (\ref{pipi}) can be discretized by evaluation on a lattice, for the purposes of numerics, it is better to discretize earlier.  For the case of $d=4$, we discretize the Hamiltonian (\ref{Hvecl}) by introducing a lattice $\theta_j = (j-1/2)\epsilon $, $j=1, \ldots, N$, where $\epsilon = \pi / N$.  (By putting the lattice at half integral points, it is easier to evaluate the behavior of the entanglement entropy with increasing $N$.)  We evaluate $\partial_\theta^2 \Phi$ using the usual second order accurate scheme: $(\Phi_{j+1} - 2 \Phi_j + \Phi_{j-1})/\epsilon^2$.  At the endpoints, we use the Dirichlet boundary condition to determine $\Phi_{N+1} = -\Phi_{N}$ and $\Phi_0 = -\Phi_1$.   

 For the case of $d=3$, we find that a grid in $u = \cos \theta$ works much better.
We consider instead
\be
\Phi_{\vec l}(\theta) = \sqrt{\sin \theta} \, \phi_{\vec l} ( \cos \theta) \ , \; \; \;
\Pi_{\vec l} (\theta) = \sqrt{\sin \theta} \, \pi_{\vec l} (\cos \theta) \ .
\ee
Note that we have now
\be
[ \phi_{\vec l} (u), \pi_{\vec l'}(u') ] = i \delta_{\vec l, \vec l'} \delta(u-u') \ .
\ee
In terms of $u$, the Hamiltonian can be written
\begin{eqnarray}
H_{\vec l} &=& \frac{1}{2R^2} \int_{-1}^1 \biggl\{ R^2 \pi_{\vec l}^2 - \phi_{\vec l} {\cal D} \phi_{\vec l} \biggr\} du 
\end{eqnarray}
where
\be
{\cal D} \phi_{\vec l} = \partial_u ((1-u^2) \partial_u \phi_{\vec l} ) - \frac{ \left(m+\frac{d-3}{2} \right)^2}{1-u^2} \phi_{\vec l} - \frac{1}{4} \phi_{\vec l} \ ,
\ee
and we dropped a total derivative.
We choose a grid with lattice points at $u_j = -1 + \left(j-\frac{1}{2} \right) \epsilon$, $j=1, \ldots, N$, and $\epsilon = 2/N$.  
We discretize the operator 
\begin{eqnarray}
\partial_u ((1-u^2) \partial_u f) &\approx& \frac{1}{\epsilon^2} \left( f_{j-1} \left(1 - \left(\frac{u_{j-1}+u_j}{2}\right)^2 \right)
+ f_{j+1} \left(1 - \left(\frac{u_{j}+u_{j+1}}{2}\right)^2 \right) + \right. \nonumber  \\
&& \left. + f_j \left( -2 + \left( \frac{u_{j-1}+u_j}{2}\right)^2 +  \left( \frac{u_{j}+u_{j+1}}{2}\right)^2 \right) \right) \ ,
\end{eqnarray}
valid at second order in $\epsilon$.  This discrete difference has the advantage that the contributions from the ghost points $u_0$ and $u_{N+1}$ vanish.
Either discretization scheme will work in $d>4$, but unfortunately because our main interest is in $d=3$ and $d=4$, we had to develop both schemes.

The most straighforward approach to checking the main result (\ref{mainresult}) is to compute $S_A(T) - S_A(0)$ using the expression (\ref{SAfreedef}), fixing $RT$ and scanning over $\theta_0$.  
This approach runs into an order of limits issue.
As discussed in ref.\ \cite{Cardy:2014jwa} in the two dimensional case, the result (\ref{mainresult}) is valid in the limit where $T$ goes to zero first.  Scanning over $ \theta_0$ leads to a cross over behavior for $\theta_0$ sufficiently large.  In the limit where $\theta_0 \to \pi$ first, the leading correction to $\delta S_A$ is given by the thermal entropy:
\be
\delta S_A =g \left( 1 + \frac{\Delta}{T R} \right) e^{-\Delta / RT} + \ldots \ .
\ee
as demonstrated in \cite{Cardy:2014jwa}.  Our numerical results for $S_A(T) - S_A(0)$ are shown in figure \ref{deltaSconstantT}.  The results exhibit precisely this  cross over behavior.  For small $\theta_0$, the agreement with the analytic result (\ref{mainresult}) is quite good (modulo the $\sin^{d-2} \theta_0$ discrepancy).  However, as $\pi - \theta$ becomes small compared to $RT$, the entanglement entropy difference looks more and more like the thermal entropy and asymptotes to it in the limit $\theta \to \pi$.  

A better numerical technique is to expand the $\coth$ functions in (\ref{phiphi}) and (\ref{pipi}), isolating the $e^{- \Delta / RT}$ dependence of $\delta S_A$ analytically.  Expanding the entanglement entropy (\ref{SAfreedef}) in the limit of small $T$, we obtain 
\be
\delta S_A = \tr \left[ \delta C_0 \cdot C_0^{-1} \cdot \log \frac{C_0+1/2}{C_0-1/2} \right] e^{-\omega_0 / T}+ \ldots\ ,
\label{SAexp}
\ee
where
\begin{eqnarray*}
\delta C_m(\theta_1, \theta_2) &\equiv&  \int_0^{\theta_0} \left[ \langle \Phi_{\vec l}(\theta_1) \Phi_{\vec l}(\theta) \rangle \delta \Pi_m(\theta, \theta_2)
+ \delta \Phi_m(\theta_1, \theta) \langle  \Pi_{\vec l}(\theta) \Pi_{\vec l}(\theta_2) \rangle \right] d \theta\ , \\
 \delta \Phi_m(\theta, \theta') &\equiv& U_m(\theta) \frac{1}{\omega_m} U_m(\theta') \ , \; \; \;
 \delta \Pi_m(\theta, \theta') \equiv U_m(\theta) \omega_m U_m(\theta') \ .
 \end{eqnarray*}
To evaluate (\ref{SAexp}), we diagonalize $C_0$, finding its left $\langle \lambda_i |$ and right eigenvectors $| \lambda_i \rangle$.  Then we insert resolutions of the identity $\Id = | \lambda_i \rangle \langle \lambda_j | / \langle \lambda_j | \lambda_i \rangle$ around $\delta C_0$. 
The thermal correction $\delta S_A$ computed in this way now agrees with the analytic calculation over essentially the whole range $0 < \theta_0 < \pi$ although once $\pi- \theta_0 \sim \epsilon$, there are some lattice effects.    See figure \ref{deltaSTfirst} for results in $d=3$ and 4.  We find similar agreement (not shown) for $d=5$ and 6.

Note that in all cases with the numerics we are able to match not the main result (\ref{mainresult}) but the main result (\ref{mainresult}) with $I_d(\theta_0)$ replaced by $I_{d-2}(\theta_0)$.  
We are confident that we have not made a $d \to d-2$ typographical error in the numerics.  The reason is that we have computed the mutual information involving two caplike regions numerically and the results agree with previous analytic computations in the literature \cite{Cardy:2013nua,Casini:2009sr}.  A naive shift $d \to d-2$ would destroy this agreement.  We describe these checks in the appendix.

\section{Discrepancies and Boundaries}
\label{sec:discrepancy}

For a manifold $M$ with boundary $\partial M$, the action for a conformally coupled scalar must be supplemented by a boundary term
(see for example \cite{Barvinsky:1995dp}):
\be
S = - \frac{1}{2} \int_M d^d x \sqrt{-g} \left[ (\partial_\mu \phi)(\partial^\mu \phi) + \xi {\mathcal R} \phi^2 \right] - \xi \int_{\partial M} d^{d-1} x \sqrt{-\gamma} \, K \phi^2 \ .
\ee
Here $K = \nabla_\mu n^\mu$ is the trace of the extrinsic curvature of $\partial M$, $n^\mu$ is a unit outward pointing normal vector to $\partial M$, and $\gamma_{\mu\nu}$ is the induced metric on $\partial M$.  
Without this boundary term, variations of the action with respect to the metric will have dependence on derivatives of the metric variation, $\delta g_{\mu\nu,\lambda}$.  This boundary term has another role; it is sufficient to preserve invariance of the action under Weyl transformations.  In the presence of this boundary term, to have a good variational principle, the usual Neumann boundary condition $n^\mu \partial_\mu \phi = 0$ is replaced by 
\be
n^\mu \partial_\mu \phi + 2 \xi K \phi= 0\ .
\label{neumann}
\ee

The boundary term poses a problem for us because the boundary $u \to \infty$ on ${\mathbb R} \times H^{d-1}$ is subtly different from the pull back of the boundary $\theta=\theta_0$ on ${\mathbb R} \times S^{d-1}$.
Away from the limit $u \to \infty$, the difference is apparent.  A constant $\theta$ slice on ${\mathbb R} \times S^{d-1}$ maps to a surface in ${\mathbb R} \times H^{d-1}$ which depends on both $u$ and $\tau$.  At $\tau = 0$, we can arrange for a constant $u$ slice to be tangent to the pull back of a constant $\theta$ slice, but away from $\tau=0$, these two surfaces do not intersect.
In the limit $u \to \infty$, the surfaces become coincident, but still their normal vectors $n_{(\theta)}^\mu$ and $n_{(u)}^\mu$ do not coincide:
\begin{eqnarray}
(n^\tau_{(u)}, n^u_{(u)}) &=& (0, 1/R) \ , \\
(n^\tau_{(\theta)}, n^u_{(\theta)}) &=& (\sinh(\tau/R) , \cosh (\tau/R) / R) \ .
\end{eqnarray}
Correspondingly, the traces of their extrinsic curvature, even at $\tau=0$, do not agree:
\be
\left. K_{(\theta)}\right|_{\tau = 0} = \frac{d-1}{R} \; ; \; \; \;
\left. K_{(u)} \right|_{\tau=0} = \frac{d-2}{R} \ .
\label{Kdiff}
\ee

Identifying the modular Hamiltonian of the region $A$ with the Hamiltonian on hyperbolic space, as we did in section \ref{sec:analytic}, required that the Euclidean partition function on ${\mathbb R} \times H^{d-1}$ be thermal with temperature $T = 1/2 \pi R$.  
On the one hand, in order for the scalar field to be at thermal equilibrium in hyperbolic space, we should choose a time independent Hamiltonian and corresponding time independent boundary $u \to \infty$.  Mapping this choice to the region $A$, the boundary condition (\ref{neumann}) will produce logarithmic singularities on $\partial A$.  In more detail, the field $\phi$ has two different possible fall-offs at large $u$, proportional to $e^{-(d-2) u/2}$ and $u \, e^{-(d-2)u/2}$.  In order to satisfy the boundary condition (\ref{neumann}) for the $u \to \infty$ boundary, 
\[
\partial_u \phi = - \frac{(d-2)^2}{2(d-1)} \phi \ ,
\]
we need to keep both behaviors, and the leading $u \, e^{-\Delta u}$ behavior will produce the logarithmic singularities on $A$.  
In contrast, if we start with the pull-back of the $\partial A$ boundary, then the boundary condition (\ref{neumann}) at $\tau=0$,
\[
\partial_u \phi = -\frac{d-2}{2} \phi \ ,
\]
 is satisfied provided we set the leading fall-off $u \, e^{-(d-2) u/2}$ to zero.  In this case, the field $\phi$ remains finite on $A$.  However, the pull-back of the $\partial A$ boundary is time dependent in hyperbolic space, leading to a time dependent Hamiltonian.  Given this time dependence,  the system is presumably not described by a thermal density matrix.
 
We have a simple remedy at hand for this difference in boundaries and boundary conditions.  We can add a counter-term to the action that uses the $u \to \infty$ boundary,
\be
S_{\rm ctr} = c \int_{\partial M} d^{d-1} x \sqrt{-\gamma} \, \phi^2 \ .
\ee
We then adjust the constant $c$ such that the boundary condition matches the boundary condition for the action that uses the pull-back of the $\partial A$ boundary, at $\tau=0$.  
  This value, $c=-\xi/R$, is set by the difference of the extrinsic curvatures (\ref{Kdiff}).  
The counter term, which is essentially minus a potential term, then adjusts the value of the modular Hamiltonian:
\be
\Delta H_{M}  = 2 \pi \xi \int_{\partial H^{d-1}} d^{d-2} x \sqrt{-\gamma} \, \phi^2 \ .
\ee
(As before, we have included a factor of $1/T$ in the definition of $H_M$.)

%
%

We now perform a change of variables to express $\Delta H_M$ in terms of an integral over $\partial A$:
\be
\Delta H_M = 2 \pi  \xi  \int_{S^{d-2}} \phi^2  \vol(S^{d-2}) \, (R   \sin \theta_0)^{d-2}\ .
\ee
For the conformally coupled scalar, the first excited state on the $S^{d-1}$ is the constant mode.  The correlation function
$\langle \phi | \phi(x)^2 | \phi \rangle$ is the classical value of $\phi(x)^2$  for this constant mode, times a factor of two because of the two possible contractions.  Using the usual relativistic normalization that includes a factor of $1/2E_\phi$, we conclude that
\be
\langle \phi | \phi(x)^2 | \phi \rangle = \frac{2}{(d-2) R^{d-2} \Vol(S^{d-1})} \ .
\ee 
(A way to check this normalization is to compute the full Hamiltonian for the conformally coupled scalar and compare with the general result (\ref{IAresult}).)
Assembling the pieces, we find that
\be
\langle \phi | \Delta H_M  | \phi \rangle  = 2 \pi  \Delta \frac{\Vol(S^{d-2})}{\Vol(S^{d-1})} \frac{\sin^{d-2} \theta_0}{(d-2)(d-1)} \ ,
\ee
which is precisely the mismatch between the calculations in sections \ref{sec:analytic} and \ref{sec:numeric}.

\section{Discussion}
\label{sec:discussion}

In the context of conformal field theory, 
we have presented some simple, general results for thermal corrections to entanglement entropy for caplike regions on spheres.
The two results to remember are the leading thermal correction to $S_A(T) - S_A(0)$, eq.\ (\ref{mainresult}), and to $S_A(T) - S_{\bar A}(T)$, eq.\ (\ref{mainresultb}).  Although we derived eq.\ (\ref{mainresultb}) from eq.\ (\ref{mainresult}), we are struck by the $d$ independence and simplicity of eq.\ (\ref{mainresultb}).  Perhaps there is another simpler derivation, perhaps one that takes as a point of departure the fact that $S_A(0) = S_{\bar A}(0)$.  

We found an interesting mismatch between our general result (\ref{mainresult}) and the particular example of a conformally coupled scalar field.  In section \ref{sec:discrepancy}, we traced the origin of this discrepancy to a boundary term in the action.  The extrinsic curvature of the $u \to \infty$ boundary in ${\mathbb R} \times H^{d-1}$ was different from the extrinsic curvature of the pull back of the $\theta = \theta_0$ boundary in ${\mathbb R} \times S^{d-1}$.  This difference in curvatures led to the fact that the Hamiltonian we used to compute the entanglement entropy in section \ref{sec:analytic} differed from ``true'' modular Hamiltonian by a boundary term, and hence to a discrepancy with the later numeric calculation of the entanglement entropy for the conformally coupled scalar.
One conclusion to draw is that in general the $\sin^{d-2} \theta_0$ dependent term in our main result, most easily extracted from the recurrence relation (\ref{recrel}), cannot be trusted.  One must first verify that the conformal field theory action lacks boundary terms that are sensitive to the difference between the $u \to \infty$ and $\theta=\theta_0$ surfaces.  The presence of such boundary terms may shift the coefficient of the $\sin^{d-2} \theta_0$ term.  

The discrepancy involving this $\sin^{d-2} \theta_0$ area law term is reminiscent of another area law scaling of the entanglement entropy with an undetermined coefficient.
Recall that the leading, zero temperature contribution to the entanglement entropy is proportional to the area of $\partial A$ \cite{Srednicki:1993im}, 
\be
S_A(0) \sim \left( \frac{R \sin \theta_0}{\epsilon} \right)^{d-2} \ ,
\label{Sredres}
\ee
where $\epsilon$ is a small distance cut-off that depends on the regularization scheme.


There have been two recent discussions of related discrepancies involving entanglement entropy and conformally coupled scalars \cite{Hung:2014npa,Lee:2014zaa}.  In these two papers, the focus is on a discrepancy between computations using the replica method and computations using the modular Hamiltonian.  The later paper \cite{Lee:2014zaa} suggests that the discrepancy arises because of boundary terms associated with the conical singularity in the replica method.  The arguments presented here appear to be similar in spirit to if different in detail from ref.\ \cite{Lee:2014zaa}.

The original motivation for this project came from an interest in the holographic result for the entanglement entropy \cite{Ryu:2006bv}.
 The holographic formula captures only the leading, linear in central charge contribution to the entanglement entropy in a large central charge limit.  
 As our corrections (\ref{mainresult}) and (\ref{mainresultb}) are independent of the central charge and subleading in this expansion, the holographic entanglement entropy formula will not duplicate them.  
 There has been recent progress in calculating subleading corrections to the holographic result.  For example, 
 a holographic calculation of the correction (\ref{mainresult}) was carried out for $d=2$ in ref.\ \cite{Barrella:2013wja}.  it would be interesting to see if there is a holographic prescription for calculating (\ref{mainresult}) or (\ref{mainresultb}) when $d>2$.  
 Note ref.\ \cite{Faulkner:2013ana} also discusses thermal corrections to entanglement entropy using a holographic dual gravity description.  They study not conformal field theories but field theories containing massive particles such that the size of the region $\ell \gg 1/m$.  They argue that in this limit the corrections should be extensive in the field theory volume.  Our results are not extensive in the volume.  However, there is no contradiction; the conformal nature of our field theory forces us to work in a different limit where $\ell \lesssim 1/m$.

Given our results for entanglement entropy, it would be interesting if the work here could be extended to include thermal corrections to the R\'enyi entropies as well.  Recall the $n$th R\'enyi entropy of region $A$ is defined to be
\be
S_n \equiv \frac{1}{1-n} \log \tr \rho_A^n \ .
\ee
In ref.\ \cite{Cardy:2014jwa}, universal thermal corrections were calculated for both the entanglement entropy and the R\'enyi entropies in $d=2$.  
Using the methods in this paper, such a calculation would naively seem to involve evaluating $n$-point functions of the stress tensor, but given the success in $d=2$, perhaps a simpler approach can be found.

\vskip 0.1in

\section*{Acknowledgments}
I would like to thank K.~Balasubramanian, F.~Benini, J.~Cardy, S.~Giombi, I.~Klebanov, K.~Jensen, N.~Lashkari, T.~Nishioka, E.~Perlmutter, S.~Pufu, M.~Ro\v{c}ek, M.~Spillane, P.~van~Nieuwenhuizen,
and especially H.~Casini  
for discussion.  This work was supported in part by the NSF under 
Grant No.\ PHY13-16617.  I thank the Sloan Foundation for partial support.

\appendix

\section{Mutual Information}

Given the fact that we needed to make the substitution $I_d(\theta) \to I_{d-2}(\theta)$ to find agreement with the numerics, one might worry that there is a bug in the numerical algorithm.  To gain confidence that the computer code is  functioning correctly, we study two limits of a particular type of mutual information.  In particular, consider the
mutual information in $d$ spacetime dimensions (at $T=0$)
\be
M_d = S_A + S_B - S_{A \cup B} \ ,
\ee
 for two regions $A$ and $B$, one with latitudes $\theta < \theta_1$ and one with latitudes $\theta > \theta_2$.  
In the limits $\theta_1, \theta_2 \ll 1$ and also $\theta_1 \approx \theta_2$, we find agreement with analytic predictions for $M_d$
by Cardy \cite{Cardy:2013nua} and by Huerta and Casini \cite{Casini:2009sr} respectively.  These predictions depend on $d$, and a naive shift $d \to d-2$ in the code would destroy the agreement.
Note that if there is a discrepancy in our calculation of the entanglement entropy proportional to the areas of 
$\partial A$ and $\partial B$, the mutual information will not be sensitive to it.  
The mutual information is designed to remove area law dependence from the entanglement entropy.

To begin, we claim that the mutual information $M_d(x)$ 
depends only on a cross ratio $x$ constructed from geometric data describing $A$ and $B$.
Having fixed $d$, that $M_d$ is a function only of $x$ follows from the facts that $M_d$ is invariant under conformal transformation and that ${\mathbb R} \times S^{d-1}$ is conformally related to Minkowski space 
(see the appendix of  \cite{Candelas:1978gf}):
\begin{eqnarray}
ds^2 &=& -dt^2 + dr^2 + r^2 d \Omega^2 \\
&=& \Omega^2 ( - d \tau^2 + d \theta^2 + \sin^2 \theta d \Omega^2) \ ,
\end{eqnarray}
where
\begin{eqnarray}
t \pm r &=& \tan \left( \frac{ \tau \pm \theta}{2} \right) \ , \\
\Omega &=& \frac{1}{2} \sec \left( \frac{ \tau + \theta}{2} \right) \sec \left( \frac{\tau - \theta}{2} \right) \ ,
\end{eqnarray}
and $d\Omega^2$ is a line element on a unit sphere.
Note that the surface $t=0$ gets mapped to $\tau = 0$, and on this surface $r = \tan( \theta/2)$.

This transformation maps the spheres $S^{d-2}$ bounding regions $A$ and $B$ at $t=0$ to concentric $S^{d-2}$ at $\tau=0$ in flat space. 
Given two spheres in flat space, we can construct only one quantity that is invariant under conformal transformation: the cross ratio.   We draw a line through the centers of the spheres.  This line will intersect one sphere at points $p_1$ and $p_2$ and the other at points $q_1$ and $q_2$.  We define the cross ratio to be
\be
x \equiv  \frac{|p_1 - p_2| |q_1 - q_2|}{ |p_1 - q_1| |p_2 - q_2|} \ .
\ee
As the mutual information is invariant under conformal transformation, $M_d$ can be a function only of 
\be
x = \frac{4 r_1 r_2}{|r^2 - (r_1-r_2)^2|} \ ,
\ee
where $r_1$ and $r_2$ are the radii of the spheres and $r$ is the distance between their centers.
In our case, $r=0$, and the cross ratio can be expressed in terms of angles as
\be
x = \frac{\sin \theta_1 \sin \theta_2}{\sin^2 \frac{\theta_1-\theta_2}{2}} \ .
\label{crangles}
\ee

\begin{figure}
 \begin{center}
  \includegraphics[width=5in]{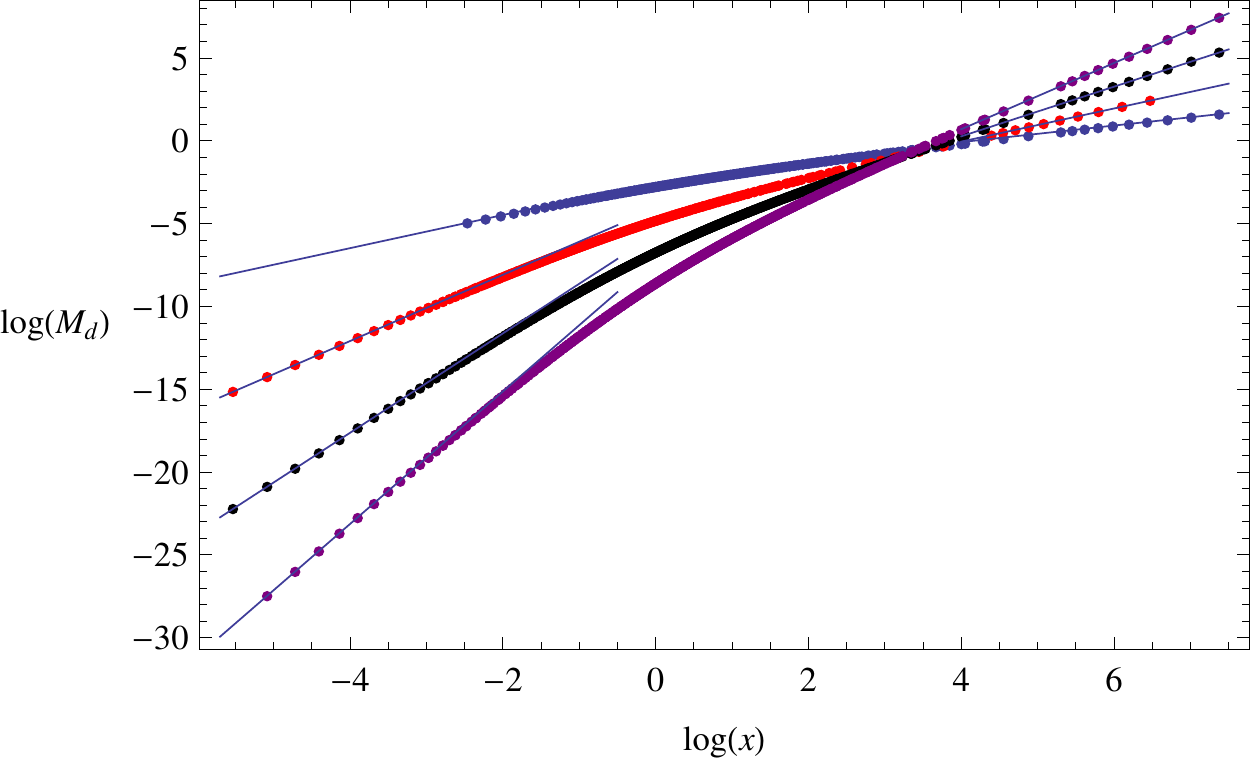}
   \end{center}
  \caption{Mutual information $M_d$ 
  for a conformally coupled scalar with two cap like regions centered around the north and south poles on $S^{d-1}$.
  The cross ratio $x$ is defined in eq.\ (\ref{crangles}).  From top to bottom on the left hand side: $d=3$, 4, 5, and 6. The order is reversed on the right hand side.  The straight lines on the right hand side are the analytic predictions by Huerta and Casini \cite{Casini:2009sr}. The straight lines on the left hand side are the analytic predictions by Cardy  \cite{Cardy:2013nua} ($d=3$, 4) or using his method ($d=5$, 6).
  Zooming in on the plot reveals that the four curves intersect at six points rather than one point.   
   } 
\label{fig:universalMI}
 \end{figure}

In terms of $x$, the two limits of $M_d$ we consider are $x \to 0$ and $x \to \infty$.  
Cardy \cite{Cardy:2013nua}, following up numerical work by Shiba \cite{Shiba:2012np}, demonstrated that $M_d(x)$ has a universal scaling behavior in the limit where $x$ becomes small.  In particular, for our conformally coupled scalar in $d$ dimensions, he argued that
\be
M_d(x) = \lambda_d \, x^{d-2} + O(x^{2(d-2)}, x^d) \ ,
\label{Mdsmallx}
\ee
In $d=3$ and $d=4$, he calculated that $\lambda_d = 1/12$ and 1/60 respectively.
We extend his computations below and argue that\footnote{%
 The sequence $1/\lambda_d$ are called the Ap\'ery numbers.  We are unsure of the significance that $1/\lambda_d$ is always an integer.
}
\be
\lambda_{d+2} =  \frac{1}{2} \frac{(d!)^2}{(2d+1)!} \ .
\label{Mdsmallxcoeff}
\ee
We find good agreement with (\ref{Mdsmallx}) and (\ref{Mdsmallxcoeff}) for $d=3$, 4, 5, and 6 (see figure \ref{fig:universalMI}
and table \ref{mytable}).
Our numerics is not sufficiently good to determine the subleading terms.

The $x \to \infty $ limit can be compared with universal behavior of the mutual information when the two regions $A$ and $B$ (in flat space) are separated by a small distance $\epsilon$:
\be
M_d \approx \kappa_d \frac{\operatorname{Area}(\partial A)}{\epsilon^{d-2}} \ .
\ee
Casini and Huerta \cite{Casini:2009sr} calculated the values of $\kappa_d$ for free bosons (see table \ref{mytable}).
We can re-express the area and $\epsilon$ in terms of the cross ratio:
\be
M_d(x) \approx \kappa_d \Vol(S^{d-2}) \left(\frac{x}{4} \right)^{(d-2)/2} \ .
\ee
We find good agreement numerically with this scaling behavior (see figure \ref{fig:universalMI} and table \ref{mytable}).

To perform the numerical calculations, we worked with a number of different grid sizes, from $N=100$ to $N=600$.  The data points were calculated by extrapolating a large $N$ limit from the finite grids assuming linear convergence in $1/N$.  
In the limit $x \to 1$, the sum over angular modes needs to be carried to large values of $m \sim O(100)$.
In the limit $x \to 0$, because of the smallness of $M_6(x)$, our accuracy was limited by machine epsilon.  Our accuracy in these limits was also limited by lattice effects and a consequent need to look at larger lattices.
The $\theta$ lattice gives better coverage at the poles and potentially better estimates of $\lambda_d$ while the $\cos \theta$ lattice gives better coverage at the equator and potentially better estimates of $\kappa_d$.  Unfortunately, the $\theta$ lattice is badly behaved in $d=3$ while the $\cos \theta$ lattice is badly behaved in $d=4$.

 \begin{table}
 \begin{center}
 \begin{tabular}{|c|c|c|c|c|}
  \hline
 &  \multicolumn{2}{|c|}{$1/\lambda_d$} & \multicolumn{2}{|c|}{$\kappa_d$}  \\
 \hline
$d$ & prediction \cite{Casini:2009sr} & fit &prediction \cite{Cardy:2013nua} & fit \\
 \hline
 3 & $12$ & $11.99$ & $3.97 \times 10^{-2}$  &  $3.85 \times 10^{-2}$ \\
 4 & $60$ & $60.08$ & $ 5.54 \times 10^{-3}$  & $5.48 \times 10^{-3}$ \\
 5 & $280$ & $280.1$ & $1.31 \times 10^{-3}$ &  $1.30 \times 10^{-3}$ \\
 6 & $1260$ & $1267$ & $4.08 \times 10^{-4}$ & $3.99 \times 10^{-4}$ 
 \\
 \hline
 \end{tabular}
 \end{center}
 \caption{
Least square fits of the coefficients $\lambda_d$ and $\kappa_d$ compared with predictions.
 \label{mytable}
 }
  \end{table}
 
 \subsection{The small $x$ limit}

We extend Cardy's calculation \cite{Cardy:2013nua} of $M_3(x)$ and $M_4(x)$ in the $x \to 0$ limit to general $d$.  
The mutual information can be extracted from an $n\to 1$ limit of the $n$th mutual R\'enyi information $M_{d,n}$.  The small $x$ limit of the R\'enyi mutual information can be calculated using the replica trick from two point functions of the scalar field on an $n$ sheeted covering of flat space branched over the origin:
\be
M_{d,n}(x) =  \frac{ n}{2(n-1)}    x^{d-2}  \left[ \sum_{j=1}^{n-1} \langle \phi_j(y) \phi_0(y) \rangle_n ^2
+ \langle {:} \phi_0(y)^2 {:} \rangle_n ^2 \right]  + O(x^{2(d-2)} ,x^d)\ ,
\label{MdCardy}
\ee
where $\phi_j(y)$ lives on the $j$th sheet.  This branched cover is $C_n \times {\mathbb R}^{d-2}$ where $C_n$ is a two dimensional cone with opening angle $2 \pi n$.  Each $2\pi$ wedge of $C_n$ corresponds to a different sheet.  Parametrizing each sheet with the coordinates $(\rho, \theta, \vec z)$, we take $y = (1,0, \vec 0)$.  

To compute two-point correlation functions of interest, we start by computing $\langle \phi (y) \phi(y') \rangle_{1/m}$ on a cone of opening angle $2\pi/m$ where we choose $y = (1,\theta,\vec 0)$ and $y' = (1,0 ,\vec 0)$.  By the method of images
\be
\langle \phi (y) \phi(y') \rangle_{1/m} = \sum_{k=0}^{m-1} \frac{1}{\left(2 \sin \frac{\theta + 2 \pi k / m}{2} \right)^{d-2} } \ .
\label{images}
\ee
The two point function $\langle \phi_j(y) \phi_0(y) \rangle_n$ can then be obtained upon replacing $m$ with $1/n$ and $\theta$ with $2\pi j$.

Eqs.\ (\ref{MdCardy}) and (\ref{images}) can be computed through careful consideration of the following two sums:
\begin{eqnarray}
S_\alpha(m,\theta) &\equiv& \sum_{k=0}^{m-1} \frac{1}{\left( 2 - 2 \cos \left(\theta + \frac{2 \pi k}{m} \right) \right)^\alpha}  \ , \\
T_\alpha(m) &\equiv& \sum_{k=1}^{m-1} \frac{1}{\left( 2 - 2 \cos \left(\frac{2 \pi k}{m} \right)\right)^\alpha} \ .
\end{eqnarray}
Note that $S_\alpha=\langle \phi(y') \phi(0) \rangle_{1/m}$ in $d = 2 \alpha+2$ dimensions.  Knowing $T_\alpha(m)$  allows one to evaluate the sum in eq.\ (\ref{MdCardy}).
We introduce the more general sums
\begin{eqnarray}
f_\alpha(m,\theta, z, \bar z) &=& \sum_{k=0}^{m-1} \frac{1}{|z - e^{i(\theta + 2 \pi k/m)}|^{2\alpha} } \ , \\
g_\alpha(m,z, \bar z) &=& \sum_{k=1}^{m-1} \frac{1}{|z - e^{2 \pi i  k/m}|^{2\alpha}} \ , 
\end{eqnarray}
where $\theta$ is real and $z$ is complex,
such that
\begin{eqnarray}
\lim_{z,\bar z \to 1} f_\alpha(m,\theta, z,\bar z) &=& S_\alpha(m,\theta) \ ; \; \; \;
\lim_{z, \bar z \to 1} g_\alpha(m,z, \bar z) = T_\alpha(m) \ .
\end{eqnarray}
There is a trivial relation between $f_\alpha$ and $g_\alpha$:
\be
g_\alpha(m, z, \bar z) = f_\alpha(m,0,z,\bar z) - \frac{1}{|z-1|^{2\alpha}} \ .
\ee

Given these definitions, we have the following recurrence relation:
\begin{eqnarray}
\frac{\partial^2 f_\alpha}{\partial z \partial \bar z} = \alpha^2 f_{\alpha+1}(m, \theta, z, \bar z) \ ,
\label{recrelf}
\end{eqnarray}
and a similar one for $g_\alpha(m, z, \bar z)$. 
The most important computation is then of $f_1(m,\theta, z, \bar z)$, for which we find, assuming $|z| > 1$,
\begin{eqnarray}
f_1(m, \theta, z, \bar z) &=& \frac{1}{|z|^2} \sum_{k=0}^{m-1} \sum_{p=0}^\infty \sum_{p'=0}^\infty z^{-p} \bar z^{-p'} e^{i (p-p') (\theta + 2 \pi k/m)} 
\nonumber \\
&=& \frac{m}{|z|^2} \left[ \sum_{p=0}^\infty \sum_{\ell=0}^\infty z^{-\ell m -p} \bar z^{-p} e^{i \ell m \theta} + c.c. - \sum_{p=0}^\infty |z|^{-2p} \right] 
\nonumber \\
&=&
\frac{m}{|z|^2-1} \left[ \frac{1}{1-e^{i m \theta} z^{-m}} + \frac{1}{1 - e^{-i m \theta} \bar z^{-m}} - 1 \right]
 \ .
\end{eqnarray}

As we are interested only in the mutual information, let's focus on the $m \to 1$ limit to keep things simple. 
First note that
\be
f_1(m, \theta,z,\bar z) = \frac{1}{|z - e^{i \theta}|^2} + O(m-1) \ .
\ee
From the recurrence relation, it immediately follows that 
\be
f_\alpha(m, \theta,z,\bar z) = \frac{1}{|z-e^{i\theta}|^{2\alpha}} + O(m-1) \ ,
\ee
and taking the limit $z \to 1$,
\be
S_\alpha(m, \theta) = \frac{1}{(2 - 2 \cos \theta)^\alpha} + O(m-1) \ .
\ee

The leading correction to the mutual information then has two contributions.  One comes from
\be
 \langle {:} \phi_0(y)^2 {:} \rangle_n ^2= \left( \lim_{\theta \to 0} \left( S_\alpha(1/n,\theta) - S_\alpha(1,\theta) \right) \right)^2 = O(n-1)^2 \ ,
\ee
and will not contribute.
The second comes from
\be
 \sum_{j=1}^{n-1} \langle \phi_j(y) \phi_0(y) \rangle_n ^2=\sum_{k=1}^{n-1} S_\alpha(1/n,2 \pi k / n)^2 = T_{2\alpha}(n) + O(n-1)^2 \ .
\ee
The leading term must be $O(n-1)$ because the sum on $k$ is empty when $n=1$.  
Note that
\be
\frac{g_1(n,z,\bar z)}{n-1} =  \left[ \frac{1}{|z-1|^2} - \frac{1}{|z|^2 -1} \left( \frac{z \log z}{(z-1)^2} + \frac{\bar z \log \bar z}{(\bar z-1)^2} \right) \right] + O(n-1) \ .
\ee
Using the recurrence relation (\ref{recrelf}), we need to extract the $(z-1)^j (\bar z -1)^j$ term in the Taylor series expansion of $g_1(n,z,\bar z)$ to determine $T_{j+1}(n)$.  For convenience, let's define $w \equiv z-1$, assume that $w$ and $\bar w$ are independent variables, and take $\bar w \ll w$.  We find that
\begin{eqnarray}
\frac{1}{|z|^2 -1} &=& \sum_{j=0}^\infty \sum_{k=0}^j  {j \choose k} (-1)^j \bar w^j w^{-k-1} \ , \\
\frac{z \log z}{(z-1)^2} &=& \frac{1}{w} + \sum_{\ell=0}^\infty \frac{(-1)^\ell}{(\ell+1)(\ell+2)} w^\ell \ .
\end{eqnarray}
Multiplying these two sums together, we see that the $w^j \bar w^j$ term in the expansion will come from $\ell = k+j+1$ and the coefficient will be
\be
\frac{T_{j+1}(n)}{n-1} + O(n-1) = \sum_{k=0}^j {j \choose k} \frac{ (-1)^k}{(k+j+2)(k+j+3)} = \frac{[(j+1)!]^2}{(2j+3)!}  \ .
\ee

Assembling the pieces, we deduce that
\be
M_{d+2} =  \lim_{n \to 1} \frac{n T_{d}(n)}{2(n-1)}  x^{d} +  O(x^{2d} ,x^{d+2}) = \frac{1}{2} \frac{(d!)^2}{(2d+1)!} x^{d} +  O(x^{2d} ,x^{d+2}) \ .
\ee
Strictly speaking, our derivation here holds for $d$ an even integer, but it holds for $d=3$ and probably holds for general odd $d$.

\end{document}